\documentclass[10pt,letterpaper]{article}
\usepackage{opex3}

\begin{document}

\title{Generating green to red light with semiconductor lasers}

\author{Gabriele Ferrari}

\address{European Laboratory for Nonlinear Spectroscopy, Istituto Nazionale Fisica Nucleare,
INFM-CNR, Polo Scientifico-Universit\`a di Firenze, 50019 Sesto Fiorentino, Italy}

\email{ferrari@lens.unifi.it}



\begin{abstract}
Diode lasers enable one to continuously cover the 730 to 1100 nm range as well as the 370 to 550
nm range by frequency doubling, but a large part of the electro-magnetic spectrum spanning from
green to red remains accessible only through expensive and unpractical optically pumped dye
lasers. Here we devise a method to multiply the frequency of optical waves by a factor 3/2 with a
conversion that is phase-coherent and highly efficient. Together with harmonic generation, it will
enable one to cover the visible spectrum with semiconductor lasers, opening new avenues in
important fields such as laser spectroscopy and optical metrology.\end{abstract}



\bibliography{Oscillazioni}

\begin{thebibliography}{10}
\bibitem{Wavemeter}
\bibinfo{title}{{\rm The lambdameter is a Coherent WaveMaster$^{\rm TM}$ with
  0.005 nm accuracy and 0.001 nm resolution. The Fabry-Perot spectrometer has a
  confocal geometry with 1.5 GHz free spectral range, a finesse of 200 at 671
  nm and it is not sensitive to 1 $\mu$m radiation.}}
\bibitem{CrystalEfficiency}
\bibinfo{title}{{\rm The asymetric intensity of the two frequency modes is due
  to the unbalanced conversion in the frequency summing crystal.}}
\bibitem{CoupledPump}
\bibinfo{title}{{\rm The measure of the pump power coupled into the cavity is
  immune to spurious effects associated with the non optimized coupling of the
  pump beam into the optical resonator, like geometric and impedence
  matching.}}
\end{thebibliography}




\section{Introduction}

Nonlinear optics is commonly used to extend the spectrum covered by lasers over unaccessible
regions \cite{dunn99}. For instance, second harmonic generation now is a well established process
applied in frequency conversion, and with continuous wave diode lasers typically it is implemented
inside resonant enhancement optical cavities \cite{zimmermann92}. Third- and up to fifth-harmonic
generation is now obtained with pulsed lasers easily accessing the UV spectral region with
familiar infrared diode-pumped solid-state lasers. The production of sub-harmonics, on the other
hand, has important applications in metrology and quantum optics. Division in 3:1 ratio is
achieved with active phase stabilization \cite{pfister96} and, more recently, dynamical signatures
of self-phase-locking for the same process were observed \cite{zondy04}. Concerning the 2:1 ratio,
both passive and active methods for the phase stabilization were applied
\cite{nabors90,mason98,feng04}. More generally, frequency downconversion with OPO's offers a
rather flexible way to access wide regions of the infrared and near-infrared spectrum, but to
generate continuous-wave and single-frequency radiation one employs single resonant OPO's, which
require multi Watts pump lasers \cite{bosenberg96}, or double resonant OPO's which, with a modest
electronic stabilization of the composing elements, show a considerably reduced threshold
\cite{nabors90,gibson99}.

We report on the first demonstration of optical frequency multiplication by a factor 3/2. We show
that our frequency multiplier, based on a multi-resonant OPO, is inherently phase coherent,
preserving the single longitudinal character of the incident field without active phase
stabilization, efficient, with a 30 \% slope efficiency and few tens of milliWatts threshold, and
stable on time scales of the order of several minutes.

\section{The 3/2 frequency multiplier}

The converter is based on an OPO where the pump, the signal, and the idler fields are all resonant
in the cavity and which is operated at frequency degeneracy making the signal and idler
frequencies to coincide. The OPO generated field has then half the frequency of the pump, and by
inserting in the cavity a nonlinear crystal for summing the pump and the OPO fields, we are able
to generate radiation at 3/2 the pump frequency. Exact degeneracy operation is obtained owing to
the double gain of the indistinguishable splitting process with respect to all the other processes
originating signal and idler photons \cite{nabors90}.

The triple resonance condition has the advantage of reducing the threshold of oscillation on the
pump intensity down to the milliWatts level \cite{martinelli01} and allows active stabilization of
the cavity length with respect to the pump frequency. On the other hand, the dispersive behavior
of the optical elements of the cavity, i.e. mirrors and nonlinear crystals, prevents one from
controlling the frequencies of the OPO generated fields independently, which has so far made
single mode operation in triply resonant OPO's hard to achieve. In our system the triply resonant
condition allows to actively stabilize the cavity length against the pumping laser, strongly
relaxing the requirements on the passive stabilization. We observed an oscillation threshold as
low as 40 mW. By introducing an independent control on the OPO frequency modes via a fine tuning
of the relative phase accumulated between the pump and OPO-generated fields over one cavity
roundtrip we achieve the simultaneous resonance of the pump and OPO fields at frequency
degeneracy.

\begin{figure} \vspace{0mm} \begin{center}
\hspace{-0mm}\includegraphics[width=0.7\textwidth,angle=0]{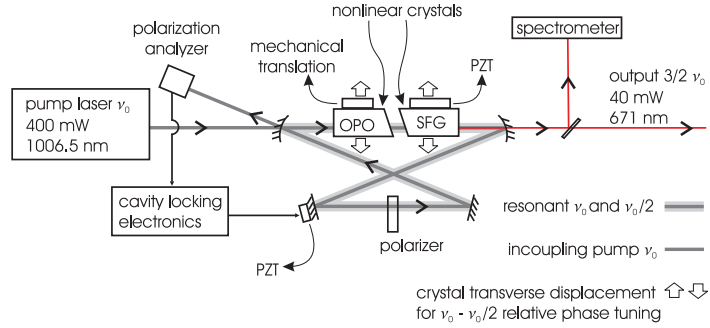}

\vspace{-0mm} \caption{\label{setup}  3/2 frequency multiplier experimental setup. A continuous
wave and single frequency pump laser delivering 400 mW of 1006.5 nm is converted into 40 mW
radiation at 671 nm. The pump laser is resonantly coupled into a cavity where 20 mm long
periodically poled KTP \cite{karlsson97} nonlinear crystals are set so to satisfy quasi
phase-matching for degenerate frequency down-conversion (OPO), and sum frequency generation
between the pump and down-converted light at 2013 nm (SFG). The wedged surfaces of the crystals
are cut at an angle of 100 mrad with respect to the crystal axis. The input (output) facet of the
OPO (SFG) crystal is at normal incidence. The two inclined surfaces facing each other are
parallel. The transverse displacement of the nonlinear crystals provides an independent control
over the cavity dispersion, insuring simultaneous resonance of the two infrared fields.}
\end{center}
\end{figure}

We demonstrate the 3/2 frequency multiplier producing radiation at 671 nm starting from a laser
source at 1006.5 nm, as schematically reported in Fig. \ref{setup}. The pump laser is composed by
a semiconductor Master-Oscillator Power-Amplifier system. The master laser is an antireflection
coated diode laser stabilized on an extended cavity in the Littrow configuration
\cite{wieman91,ricci95} delivering 30 mW at 1006.5 nm on a single longitudinal mode with less than
500 kHz linewidth. This laser is then amplified to 400 mW preserving its spectral properties
through a semiconductor tapered amplifier \cite{nyman06}. The pump radiation is coupled into an
optical cavity composed by highly reflective mirrors at 1006.5 nm and 2013 nm, and highly
transmitting at 671 nm \cite{Mirrors-671-1006.5-2013}. The input mirror has a 10 \% transmission
at 1006.5 nm in order to maximize the coupling of the pump field into the cavity under resonance.
One of the folding mirrors is mounted on a piezoelectric transducer (PZT) to actively stabilize
the cavity length to the pump field resonance. To this purpose the error signal is provided by the
polarization analysis of the reflected pump \cite{haensch80}, and by inserting into the cavity a
vertical polarizer.

The independent control on the OPO cavity frequency modes under pump resonance conditions is
obtained by cutting the crystals with a wedged shape \cite{imeshev98} (see Fig. \ref{setup}).
Displacing the crystals along the direction of the wedge enables one to change the optical path in
the crystal and, due to the dispersion, it allows a fine tuning of the OPO resonance modes while
keeping the cavity resonant with the pump field. The two nonlinear crystals are 20 mm long,
2$\times$1 mm$^2$ cross section, periodically poled KTP \cite{karlsson97} that insure
quasi-phase-matching for linearly and identically polarized fields. The OPO and SFG crystals have
a poling period of $\Lambda_{\rm OPO}$ = 38 and $\Lambda_{\rm SFG}$ = 19.5 $\mu$m respectively,
and they are identically cut in an asymmetric way such that one surface is at normal incidence,
while the other has an angle $\phi$ of 100 mrad with respect to the crystal axis. In the resonator
the crystals have the wedged side facing and parallel such that the optical axis coincide. This
configuration, while it allows to control the relative phase between the pump and OPO fields, it
insures a negligible deviation of the beam propagation at different wavelengths, and hence
simultaneous resonance of the pump and degenerate OPO fields. To reach the double resonance and
frequency degenerate condition, we observe a 400 $\mu$m periodicity on the crystal transverse
position. This is consistent with the calculated periodicity $\Lambda_{\rm OPO}/\phi$ = 380
$\mu$m. The crystal surfaces are all anti-reflection coated such that the reflectivity per surface
is 0.1\% at 1006.5 nm and 2013 nm, and 0.3\% at 671 nm.

\begin{figure} \vspace{0mm} \begin{center}
\hspace{-0mm}\includegraphics[width=0.8\textwidth,angle=0]{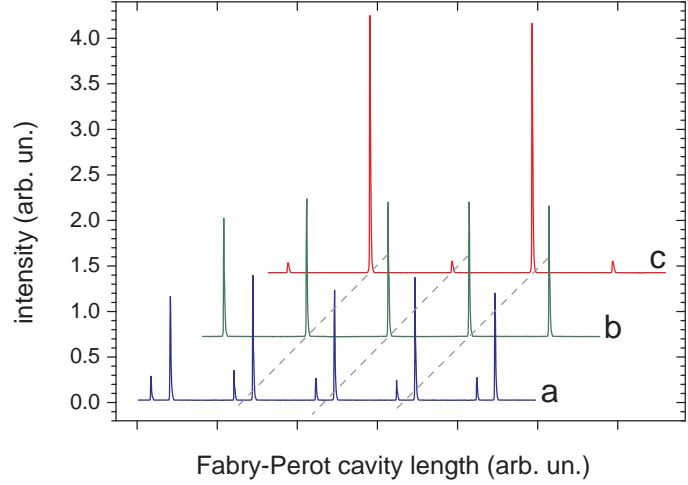}

\vspace{-0mm} \caption{\label{SpettriFP} Transmission spectra of the frequency multiplied light
through a confocal Fabry-Perot (FP) spectrum analyzer. Displacing the nonlinear crystals
transversally we tune the cavity dispersion in order to impose single frequency emission (b), or
multi mode emission (a). c) The Gaussian beam profile of the 3/2 frequency multiplied output is
verified by coupling the single frequency radiation mainly into the fundamental transverse mode of
the FP cavity, which results in doubling the spacing among the resonance peaks \cite{SpectrumFP}.
}
\end{center}
\end{figure}

\section{Spectral properties and conversion efficiency}

\begin{figure} \vspace{0mm} \begin{center}
\hspace{-0mm}\includegraphics[width=0.8\textwidth,angle=0]{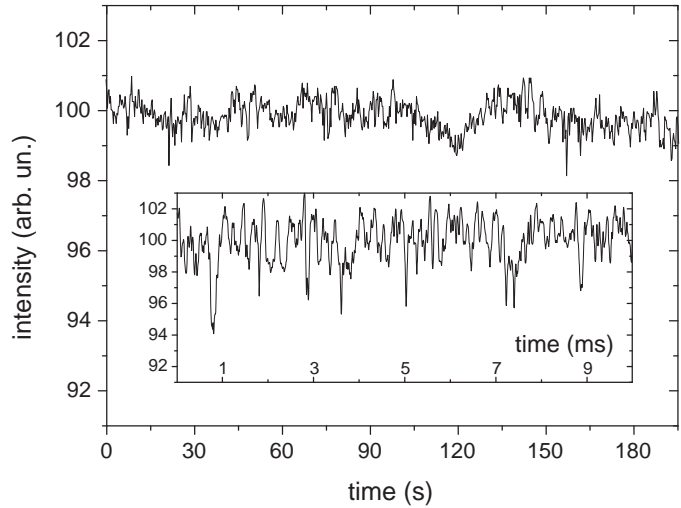}

\vspace{-0mm} \caption{\label{StabilitaAmpiezza} Frequency multiplier amplitude stability on 3
minutes and 10 ms (inset) time timescale under single longitudinal mode emission. The measured RMS
amplitude noise at full power is 1.4 \% on 50 kHz bandwidth. Under multi longitudinal mode
operation the amplitude stability does not not change qualitatively when measuring on the same
bandwidth.}
\end{center}
\end{figure}

The spectral properties of the generated red light are analyzed both with a lambda-meter for the
rough wavelength determination, and a confocal Fabry-Perot spectrometer (FP) to check the single
longitudinal mode operation \cite{Wavemeter}. As expected, the spectrum of the generated red light
depends on cavity dispersion. When we change the transverse position of the crystal by tens of
microns we are able to switch between single frequency emission at the expected value and multi
frequency emission with central wavelength displaced as much as 0.08 nm from 671 nm, with a
simultaneous reduction in the output power. Figure \ref{SpettriFP} reports the typical spectra
from the Fabry-Perot analyzer when the OPO operates close to the degenerate point. Depending on
the transverse position of the crystals, the system emits single (spectrum \ref{SpettriFP}b) or
multi (\ref{SpettriFP}a) longitudinal mode radiation with a stability of the order of minutes. In
the multi-longitudinal mode operation, energy conservation results in the symmetric positioning of
the frequency components with respect to the degenerate mode \cite{CrystalEfficiency}. The spatial
mode of the red light has a nearly Gaussian profile. As a check we carefully aligned the FP
analyzer in order to discern the even and odds transverse modes of the cavity \cite{SpectrumFP}.
As reported in Fig. \ref{SpettriFP}c, we can couple 97 \% of the power into the even transverse
modes, indicating that at least 94 \% of the generated power is in the fundamental transverse
mode. While multi-longitudinal mode operation is stable on hours, when the converter emits single
frequency radiation it proves to be stable on timescale of order of several minutes. Such a
stability requires no active stabilization of the crystals position. Figure
\ref{StabilitaAmpiezza} depicts the amplitude of the generated red light when the frequency
multiplier works in single longitudinal mode. The measured amplitude noise is 1.4 \% RMS on a 50
kHz bandwidth. \\
The single longitudinal mode emission proves that the OPO works at frequency degeneracy, and it is
known that for type-I phase matching (as provided by the periodically poled crystals) in frequency
degenerate OPO's the pump and downconverted fields are phase locked and that they may exhibit
$\pi$ phase jumps \cite{nabors90}. On the other hand the phase of the fundamental field in the
cavity is locked to that of the incident beam because of the pump-cavity resonance condition.
Since the frequency sum process should not add any relevant phase noise, we have an evidence that
the 3/2 multiplication process is phase coherent. A comparison with independently generated phase
coherent fields will allow a thorough characterization of the stability of the phase transfer
\cite{stenger02}.

\begin{figure}\vspace{0mm} \begin{center}
\hspace{-0mm}\includegraphics[width=0.8\textwidth,angle=0]{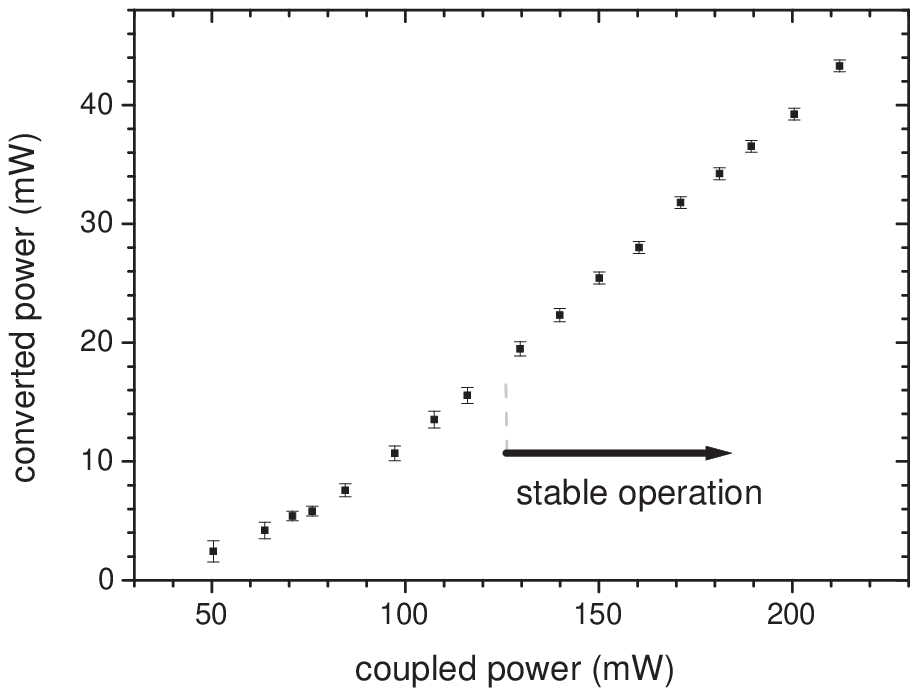}

\vspace{-0mm} \caption{\label{Efficiency} Extracted power at 671 nm as function of the pump power
coupled into the cavity. The vertical gray line indicates the threshold value for a stable single
frequency operation of the converter. The error bars correspond to the RMS amplitude noise.}
\end{center}
\end{figure}

We determine the conversion efficiency by varying the pump power and measuring the generated power
in the red as a function of the IR power coupled into the cavity \cite{CoupledPump}. We observe a
threshold for OPO oscillation smaller than 50 mW and obtain a 30 \% incremental efficiency above
150 mW pump power coupled into the cavity (see Fig. \ref{Efficiency}). Reducing the intensity of
the pump below 2/3 of the full power raises the amplitude noise in the output, and makes the
system more critical to operate on a single longitudinal mode. Such a degrading can be overcome by
using a different geometry optimized for lower pump levels, with better focussing of the cavity
mode on the nonlinear crystals, and choosing different crystals with higher nonlinear
polarizability \cite{martinelli01}.

The wavelength tunability of the source can be limited either by the tunability of the fundamental
laser, or by that of the 3/2 frequency multiplier. Typically anti-reflection coated infrared
semiconductor lasers have a tunability of few percent in wavelength, and in our case the laser can
emit from 990 nm to 1040 nm. Concerning the multiplier, the nonlinear crystals can be temperature
tuned to satisfy quasi-phase-matching at different wavelengths. With our crystals, to generate
radiation at 670 nm, one nm shorter wavelength, we have to tune master laser to 1005 nm, cool the
OPO crystal by 5 Celsius, and cool the SFG crystal by 20. With a given choice of grating periods,
a reasonable temperature tunability of the multiplier is 0.5 \% in wavelength. This can be
extended, without loss of efficiency, to the full 5 \% tunability of the pump by using
multichannel periodically poled crystals \cite{myers95bis}, which include the 10 grating periods
necessary to access the relevant wavelength intervals. The mirrors of the cavity have a flat
response beyond the window accessible through the master laser.

\section{Conclusion}

To summarize, we demonstrated for the first time a scheme to multiply the frequency of continuous
wave optical radiation by a factor 3/2 and preserving its single frequency character. The
frequency multiplier is phase coherent, has a high conversion efficiency, is very stable, and its
stability on hours timescale could easily be achieved by optimizing the design of the
opto-mechanical apparatus. Employing existing technology, the scheme will easily find applications
in many disciplines requiring laser in the green to red spectral interval, such a spectroscopy.
Together with integer harmonic generation, 3/2 frequency multiplication will allow to access the
complete visible spectrum via harmonic generation of semiconductor lasers. It makes possible to
establish phase coherent links among spectral regions distant 2/3 of an octave \cite{ramond02},
and it may considerably simplify the realization of RGB laser systems.\\
It is worth noting that the frequency multiplier also acts as a parity discriminator on the pump
resonant mode. In fact, neglecting cavity dispersion, the frequency degenerate and resonant down
conversion can take place only when the pump is resonant in the cavity with an even number of
modes. This is confirmed by the single frequency emission of the converter with a twofold
periodicity when stepping the cavity length between adjacent pump resonances.

\section*{Acknowledgment}

We thank M. Artoni, G. Oppo and N. Poli for a critical reading of the manuscript, R. Ballerini, M.
De Pas, M. Giuntini and A. Hajeb for technical assistance. We are indebted with G.M. Tino, R.
Grimm, F. Schreck and Laser \& Electro-Optic Solutions for the general support and the loan of
parts of the apparatus. We also acknowledge stimulating discussions with C. Salomon. This work was
supported by EU under contract RII3-CT-2003-506350, and Ente Cassa di Risparmio di Firenze.

\end{document}